\documentclass[prl,twocolumn,showpacs,superscriptaddress,amsmath,amssymb]{revtex4}

\usepackage{graphicx}%
\usepackage{dcolumn}%

\begin{document}

\title{Circular dichroism in angle-resolved photoemission spectra of under- and overdoped Pb-Bi2212.}

\author{S. V. Borisenko}
\affiliation{Institute for Solid State Research, IFW-Dresden, P.O.Box 270116, D-01171 Dresden, Germany}

\author{A. A. Kordyuk}
\affiliation{Institute for Solid State Research, IFW-Dresden, P.O.Box 270116, D-01171 Dresden, Germany}
\affiliation{Institute of Metal Physics of National Academy of Sciences of Ukraine, 03142 Kyiv, Ukraine}

\author{A. Koitzsch}
\author{T. K. Kim}
\author{K. Nenkov}
\author{M. Knupfer}
\author{J. Fink}
\affiliation{Institute for Solid State Research, IFW-Dresden, P.O.Box 270116, D-01171 Dresden, Germany}

\author{C. Grazioli}
\author{S. Turchini}
\affiliation{Istituto di Struttura della Materia, Consiglio Nazionale delle Ricerche, Area Science Park, I-34012 Trieste, Italy}

\author{H. Berger}
\affiliation{Institute of Physics of Complex Matter, EPFL, CH-1015 Lausanne, Switzerland}

\date{\today}
\begin{abstract}
We use angle-resolved photoemission with circularly polarized excitation to demonstrate that in the 5x1 superstructure-free Pb-Bi2212 material there are no signatures of time-reversal symmetry breaking in the sense of the criteria developed earlier (Kaminski et al. Nature {\bf 416}, 610 (2002)). In addition to the existing technique, we suggest and apply an independent experimental approach  to prove the absence of the effect in the studied compounds. The dichroic signal retains reflection antisymmetry as a function of temperature and doping and in all mirror planes, precisely defined by the experimental dispersion at low energies. The obtained results demonstrate that the signatures of time-reversal symmetry violation in pristine Bi2212, as determined by ARPES, are not a universal feature of all cuprate superconductors.

\end{abstract}

\maketitle
The variety of specific points, lines and regions in the "normal state" part of the phase diagram of the high-temperature superconductors (HTSC) clearly demonstrates not only its complexity but also the absence of its detailed understanding \cite{PhaseDiagram}. It is therefore important to realize which of them are really universal boundaries of particular phases and which just designate intermediate states with properties defined by the proximity to the well established phases such as, for instance, superconductivity. A recent angle-resolved photoemission (ARPES) study \cite{Kaminski} found evidence of time-reversal symmetry breaking below the so called T*-line, implying the existence of a well defined phase transition in underdoped cuprates. However, in the same set of experiments it was shown that the violation of the time-reversal symmetry is not peculiar only to the pseudogap regime but persists well in the superconducting state. This latter observation suggests that the detected symmetry breaking could, in principle, originate from the superconducting state and then be the primary cause of the effect in the pseudogap regime. This uncertainty calls for further experimental investigations and can be clarified if the superconducting state of the overdoped samples is studied in the same manner. Moreover, the importance of the issue appeals to the confirmation of already existing data since the observation of the effect is an extremely demanding experiment \cite{Kaminski,Varma} in which a number of artifacts should be ruled out before one can state that exactly the time-reversal symmetry breaking is responsible for the non-vanishing dichroism in the mirror plane. It was already suggested \cite{Armitage} that the $\approx3\%$ asymmetry effect  observed in the underdoped samples can be explained by the changes of the well known incommensurate modulation (reported in Ref.\onlinecite{Kaminski} to be also of the order of $3\%$) as a function of temperature. Therefore, analogous experiments carried out on the systems with reduced interference of the temperature-sensitive structural modifications together with the developement of an improved experimental methodology aiming at more precise and reliable investigation of circular dichroism effects in low energy photoemission would be of special interest today.

In this Letter we present the results of the ARPES investigation of the (Pb,Bi)$_2$Sr$_2$CaCu$_2$O$_{8+\delta}$ (Pb-Bi2212) cuprates known to have no 5x1 superstructure. In addition to the temperature dependent analysis, we consider  the symmetry of the dichroic signal (details of its momentum distribution) with respect to the momentum distribution of the total photocurrent, calculated as a sum of the spectra taken with right- and left-hand circularly polarized excitation and therefore equivalent to the photocurrent measured using unpolarized light. Application of the method to the Pb-Bi2212 compounds demonstrates that the reflection antisymmetry of the dichroic signal with respect to the mirror planes remains unsensitive to both temperature and doping level.

The experiments were performed at the 4.2R beamline "Circular Polarization" of the ELETTRA storage ring using approximately 90$\%$ circularly polarized (CP) light from the elliptical wiggler-undulator. Spectra were collected in the angle-multiplexing mode of the SCIENTA SES-100 electron-energy analyzer. The overall average resolution in ({\bf k}, $\omega$)-space was set to 0.01 {\AA}$^{-1}$ x 0.02 {\AA}$^{-1}$ x 40 meV. An essential advantage of this experimental setup is that no mechanical movement is involved in the process of switching the helicity of the incoming radiation. Only the direction of the current in the coils of the wiggler-undulator needs to be reversed which takes approximately 30 seconds. This enables the successive recording of the spectra using the light of both polarizations with the other experimental parameters remaining unchanged. Direct imaging of the beam spot (typical linear size $\sim 300 \mu$m) on the sample surface using the transmission mode of the same electron-energy analyzer has demonstrated its perfect ($< 10 \mu$m) spatial stability with respect to multiple switching of the helicity \cite{spot}. We note here that such experimental conditions are obviously more favorable for the dichroism studies than those reported in Ref.2, where not only the polarizer is rotated resulting in the residual beam movement but also the experimental chamber needs to be adjusted every time the helicity is changed.
\begin{figure}[t!]
\includegraphics[width=8.47cm]{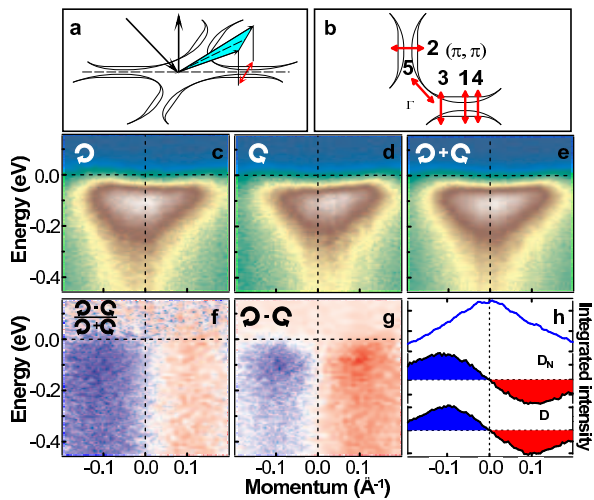}%
\caption{\label{First} a) Schematic layout of the experiment. b) Cuts in {\bf k}-space corresponding to the recorded spectra. c) and d) EDMs taken using the light of positive (c) and negative (d) helicity. e) Sum of EDMs shown in panels (c) and (d). f) the normalized difference. g) the difference. h) Integrated intensities corresponding to panels (e), (f) and (g). For details see text.} 
\end{figure}
High-quality single crystals of 5x1-superstructure-free, underdoped (T$_{c}$=77 K) and overdoped (T$_{c}$=70 K) Pb-Bi2212 were mounted on the three-axis stepper motor driven cryo-manipulator allowing a precise (0.1$^{\circ}$) positioning of the sample with respect to the analyzer's entrance slit. Alignment of the crystals was done by recording characteristic spectra with pronounced {\bf k}-dependence. The stability of the sample orientation was further controlled by a digital camera with a sensitivity to the relative movement of the sample of the order of 0.1$^{\circ}$. The excitation energy was chosen to be h$\nu$=50 eV for two reasons: (i) the emission from the antibonding band is strongly enhanced near the ($\pi$, 0)-point in comparison with that from the bonding band thus effectively reducing the number of features in the spectra \cite{KordPRL02}; (ii) the (3$\pi$, 0)-point becomes accessible in normal incidence geometry at an emission angle of 45$^{\circ}$. Intensity variations of the synchrotron radiation were controlled by continuously monitoring the ring current. If jumps of more than 1$\%$ were observed during the data acquisition, measurements have been repeated. The data were collected at 300 K, 100 K and 30 K.

The basic idea of our approach is the same as was previously suggested and experimentally tested \cite{Kaminski,Varma} but there are important differences which we discuss in the next paragraph. According to the proposed criterion one needs to control the value of the dichroic signal corresponding to the emission within a mirror plane. If this signal is zero, the time-reversal symmetry is preserved, if not - it is broken. At that, vectors of incidence, emission and normal to the sample surface should lie in this mirror plane. Such significant modification (an earlier proposal could be found in Ref. \onlinecite{Varma_old}) of the criterion was required because of the strong dichroism observed in the case when the experimental geometry posesses a "handedness", i.e. when at least one of the three mentioned vectors is not in the mirror plane \cite{remark}. Regardless of the origin of such an effect, though argued to be geometric in Ref. \onlinecite{Varma}, it can be used to track down the temperature dependence of the intrinsic dichroism (if any). The geometric effect is odd with respect to the reflection in a given mirror plane and an intrinsic dichroism, expected to be even, should then result in an effective "rotation" of this mirror plane upon entering the pseudogap regime of underdoped samples, as was observed for pure Bi2212 in Ref. \onlinecite{Kaminski}, where the corresponding angle was estimated to be 2.3$^{\circ}$. 

The essentials of our approach to test the time-reversal invariance of the electronic states in Pb-Bi2212 are depicted in Fig.\ref{First}. As shown in Fig.\ref{First}(a) we record the photoemission intensity simultaneously for {\bf k}-vectors from the cuts crossing the mirror plane at a right angle (red arrows). Resulting Energy Distribution Maps (EDM) taken at room temperature using the right- ($I^{+}$) and left-hand ($I^{-}$) CP light  are shown in Fig.\ref{First}(c) and (d) respectively. One easily notes an asymmetric character of both distributions with respect to the zero {\bf k}-values representing the mirror plane. Panels (f) and (g) show normalized, $(I^{+}-I^{-})/(I^{+}+I^{-})$, and simple, $I^{+}-I^{-}$, differences of the EDMs shown above. Here one sees how the dichroic signal is distributed as a function of energy and momentum. Its essentially homogeneous character (Fig.\ref{First}f) seems to be in agreement with its geometric origin. Fig.\ref{First}h concludes the presentation of the dataset condensing it into the three curves which are the main basis of our data analysis - (i) momentum distribution curves (MDC) corresponding to the total photocurrent (one example is shown as a solid blue line), (ii) integrated normalized difference ($\bf D_{N}$) and (iii) integrated simple difference ({\bf D}). MDCs are needed to precisely determine the {\bf k}-location of the mirror plane. The other two curves were obtained by integration within the energy interval of -450 to 100 meV of the corresponding distributions from the panels (f) and (g). This example clearly shows that the dichroic signal is zero in the mirror plane. 
\begin{figure}[t!]
\includegraphics[width=8.7 cm]{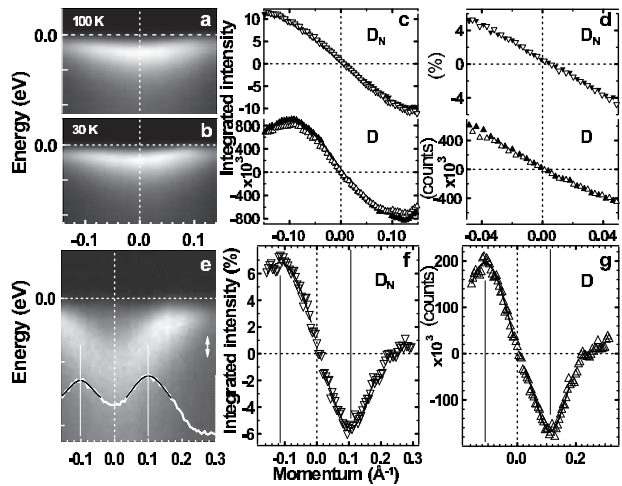}%
\caption{\label{Second} Overdoped sample. a), b) EDMs of the total intensity taken along the cut 4 (see Fig.1b) at 100 K and 30 K. c) and d) Corresponding dichroic signals. e) EDM taken along the cut 3  in the superconducting state and MDC (white line, obtained by the integration of the spectral weight within the energy range shown by the double-headed arrow) used to determine the origin of the momentum scale. f) and g) Corresponding dichroic signals with pronounced {\bf k}-dependence which allows a self-check procedure.}
\end{figure}

We stress two important points here. We {\it first} determine the position of the mirror plane with the precision of 0.002 \AA$^{-1}$ from considering MDCs of the total distribution (Fig.\ref{First}e) and then plot $\bf D_{N}$ and {\bf D} in these coordinates. In other words, we believe that the best accuracy in alignment could be achieved using "internal" reference points, such as experimental dispersion derived from the MDCs maxima of the total spectral weight. It is worth to note, that these reference points are obtained in static conditions, i.e. without any movement of the sample or the analyzer, which is ruled out when using a smaller detector at lower photon energies as in Ref.2, where neither the procedure itself nor the accuracy of the momentum scale determination was reported.

The second point is that now, when we have recorded the signal in a relatively wide momentum interval, one easily notices the non-monotonic character of both $\bf D_{N}(k)$ and {\bf D(k)}-dependencies (Fig.\ref{First} h). Locations of the extrema as well as their absolute intensities could be determined with a high precision using a fitting procedure and are symmetric with respect to the origin when the sample is properly aligned. We have systematically studied the line shape of $\bf D_{N}$ and {\bf D} as a function of misalignments of different types \cite{Boris}. Both turned out to be extremely sensitive: not only the curves did not pass through the origin imitating the presence of the intrinsic dichroism, but at the same time their maxima and minima were no longer symmetric with respect to zero. This observation opens up an additional possibility to check the reflection invariance of the dichroic signal. For instance, if some sort of a misalignment resulting in a non-zero value in the mirror plane is compensated by the intrinsic dichroism expected as a result of the time-reversal symmetry violation, one can identify such unlikely case of "accidental zero" watching the locations and absolute intensities of the extrema of $\bf D_{N}$ and  {\bf D}-curves.

Another important advantage of recording the wide-range dichroic signal follows from the remarkable property of the $D_{N}=(I^{+}-I^{-})/(I^{+}+I^{-})$ function. Its lineshape is practically independent on the intensity ratio of the right- and left-hand circularly polarized photon flux which has defined the accuracy of the experiments in Ref. \onlinecite{Kaminski}. Indeed, if $\alpha$ is a factor to account for the different flux of the photons of opposite helicities, for instance, due to the changes of the ring current, then
\begin{equation}
D_{N}'= \frac{I^{+}-\alpha I^{-}}{I^{+}+\alpha I^{-}}= \frac{D_{N} + 1 - \alpha + \alpha D_{N}}{D_{N} + 1 + \alpha - \alpha D_{N}}.
\end{equation}
One can expand $D_{N}'$ as:
\begin{equation}
D_{N}'= \frac{1 - \alpha}{1 + \alpha} + \frac{4 \alpha}{(1 + \alpha)^2} D_{N} + \frac{4 \alpha (\alpha - 1)}{(1 + \alpha)^3}D_{N}^2 + ...
\end{equation}
where, since $D_{N}$ is usually very small and $\alpha$ is of order of 1, only the two first terms are essential. It is easy to see that deviation of $\alpha$ from unity only rescales and shifts the $D_{N}$-function along the vertical axis leaving the opportunity to define {\bf k}-locations of the extrema with the same precision. From this observation it follows that one can use the $D_{N}$ line shape alone to search for the intrinsic dichroism unless the expected effect is of the same magnitude and sign for all k-vectors along the studied cut which would contradict theoretical prediction anticipating a maximum at the ($\pi$, 0)-point \cite{Varma}.

Now we apply our approach to check whether there is an effect in the superconducting state of the overdoped sample - the issue which was not addressed before. First we compare in Fig. \ref{Second} the total intensity distributions together with the $\bf D_{N}$ and {\bf D} curves measured at 100 and 30 K. In this case we observe antisymmetric behaviour of both $\bf D_{N}$ and {\bf D}, and it is not sensitive to the temperature (Fig. \ref{Second}c). In the panel (d) we replot the data in more detailed momentum scale, comparable with the one used for the presentation of the data in Ref. \onlinecite{Kaminski}. As is seen, the accuracy of our experiment is better and allows to make a conclusion about the absence of the effect of the mirror plane "rotation" in the overdoped sample upon entering the superconducting state.

Fig.\ref{Second} (e-g) illustrates that the additional condition described above is satisfied. Data are taken now along the cut closer to $\Gamma$-point which results in more clearly pronounced extrema of $\bf D_{N}$, momentum location of which we determine by fitting these parts with gaussians. The left maximum resides at $-0.115\pm 0.003$ \AA$^{-1}$ whereas the right minimum is at $0.113\pm 0.001$ \AA$^{-1}$. 
\begin{figure}[t!]
\includegraphics[width=8.47cm]{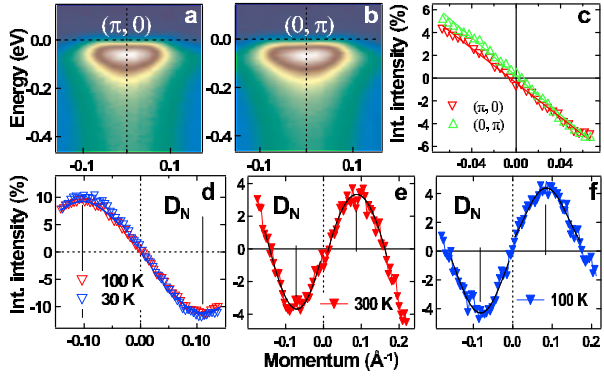}%
\caption{\label{Third} a) and b) EDMs of the total photocurrent along the cuts through the ($\pi$, 0)- and (0, $\pi$)-points (cuts 1 and 2) taken at 100K and c) correspondent $\bf D_{N}$ - curves together with linear fits. d) $\bf D_{N}$ - curves for the cut 3 measured at 100K and 30 K. e) and f) Normal incidence $\bf D_{N}$ - curves taken at 300K and 100K. For details see text.} 
\end{figure}

The data recorded on underdoped \cite{gap} samples are presented in Fig. \ref{Third}. The first row shows EDM's of total intensity together with $\bf D_{N}$-curves recorded perpendicular to the two mirror planes $\Gamma$ - ($\pi$, 0) and $\Gamma$ - (0, $\pi$) in the pseudogap state. As seen from the panel (c) the dichroism in the mirror plane is negligible. At zero momentum linear fits cross the vertical axis at 0.16$\%$ and -0.41$\%$, respectively, which is within the error bars ($<0.5\%$). These values are even much smaller than the dichroism value of $\sim 4\%$ reported in Ref. \onlinecite{Kaminski} for the pseudogap regime. In panel (d) we compare dichroic signals recorded along the cut 3 (Fig.\ref{First}, b) measured at 100 K and 30 K and again, as follows applying both criteria, there is no visible deviation from the zero. Panels (e) and (f) represent the data taken in normal incidence geometry. 
Due to the relatively large emission angle (45$^\circ$), reduced "handedness" of the experiment (because of the coplanarity of three aforementioned vectors) and radial matrix element effects, the signal is weaker and more problematic for the application of the absolut intensity criterion. Linear fits in the vicinity of the zeroth momentum cross the vertical axis at -0.63$\%$ and 0.18$\%$ respectively which is slightly more than experimental uncertainty ($\sim 0.5\%$) for the room temperature curve. This is exactly the case when one can apply additionally the line shape criterion dicussed above. Fitting the extrema of the normal incidence curves measured at 300 and 100 K we find that the "zero" is shifted by less than 0.009 \AA$^{-1}$ at room temperature and by less than 0.004 \AA$^{-1}$ in the pseudogap regime implying that the reason for the small dichroism in the mirror plane is the residual misalignments. Qualitatively similar picture was observed for other underdoped samples (results are not shown).

\begin{figure}[t!]
\includegraphics[width=8.47cm]{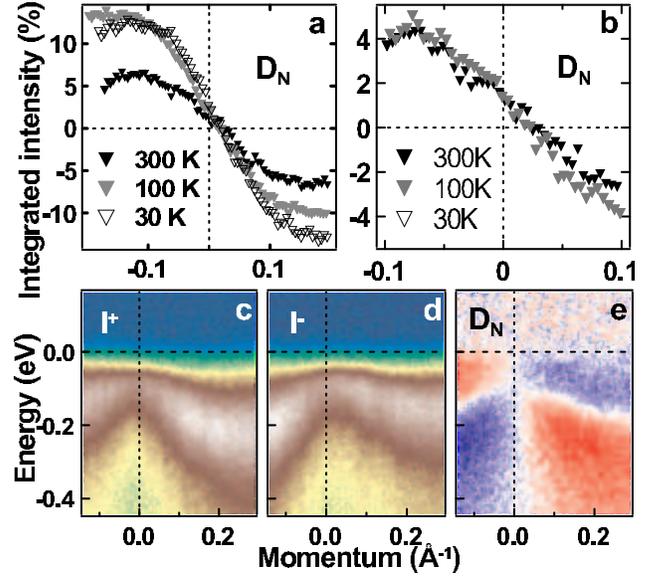}%
\caption{\label{Fourth} a) and b) Dichroic signals demonstrating the absence of rotation of $\Gamma$ - ($\pi$, $\pi$) mirror plane upon cooling the sample in overdoped and underdoped sample, respectively. c), d) and e) Plus, minus and normalized difference EDMs taken along the cut 5 (see Fig.1b). Note the different sign of the dichroism for the bonding and antibonding components.} 
\end{figure}

When discussing the time-reversal invariance it is instructive to control the dichroism behavior in the vicinity of all mirror planes of the crystal. The knowledge of the {\bf k}-distribution of the dichroism in the whole Brillouin zone when time-reversal symmetry is broken can help to identify important characteristics of the phenomenon which causes such violation, for instance, particular pattern of the circulating currents. In Fig. \ref{Fourth} we show the results for the $\Gamma$ - ($\pi$, $\pi$) mirror plane. As in the case of normal incidence, the dichroic signal is weak but the reason now is completely different. As one easily sees in the panels (c)-(e), the dichroism has opposite sign for bonding and antibonding bands. This observation is also in favor of our choice of the excitation energy for the near-($\pi$, 0) experiments - the total dichroism would be much weaker if we use, e.g., 22 eV photons resulting in a comparable contributions of bonding and antibonding bands to the total spectral weight \cite{KordPRL02}. We also point out that the alignment of the sample is quite a challenge here due to the specific momentum distribution along this cut (see Fig. \ref{Fourth} c, d). Nevertheless, as temperature dependent measurements show, there is no evidence for dramatic changes in the dichroism in under- and overdoped samples.  We have also found no evidence for the breaking of the reflection symmetry in the $\Gamma$ - ($\pi$, -$\pi$) plane (results are not shown) required by the pattern of currents suggested in Ref. \onlinecite{Varma}.

The accuracy of our experiments could be estimated in two ways. We distinguish between the temperature dependent measurements where error bars are given by the size of the relative shift of the $\bf D_{N}$-curve and measurements of a single $\bf D_{N}$-curve at a given temperature where the accuracy is defined by the absolute {\bf k}-value at which the dichroism is zero or by the absoulute dichroism value at {\bf k}=0. In both cases typical values are of the order of $\pm 0.004$ \AA$^{-1}$, which in terms of the dichroism is $\pm 0.3 \%$ or in terms of the degree of the mirror plane "rotation" would be $\pm 0.3^{\circ}$. Note, that our additional criterion allows to identify and account for the major source of the errors - misalignments.  In some particular cases (e.g. Fig. \ref{Third} e, f) the quality of the gaussian fits defines the precision of the experiment ($\pm 0.005$ \AA$^{-1}$). 

Our attempts to study pure Bi2212 single crystals led us to the following conclusions. Because of the 5:1 superstructure present in Pb-free Bi2212 samples, the dichroic signal measured in the ($\pi$, 0)-point already at the room temperature is not equal to zero. According to our measurements it is $\sim3.7\%$ and the $\bf D_{N}$ line crosses the momentum axis away from the ($\pi$, 0)-point which is easy to understand since the additional dichroism related to the two diffraction replicas crossing the Fermi level along this cut is not compensated. This is because diffraction replicas originate from the different Brillouin zones where the corresponding photoemission signal is widely known to be different due to the matrix element effects. If one, however, starts to measure the evolution of the dichroism in the point where it is zero at room temperature, i.e. not in the ($\pi$, 0)-point, it is not surprising that at lower temperatures the dichroism can become non-zero. Moreover, this effect will be doping dependent since in the overdoped samples diffraction replicas are less important crossing the Fermi level further away from the ($\pi$, 0)-point than in underdoped samples.

The obtained results do not rigorously disprove the existance of the time-reversal symmetry breaking phase in the pseudogap part of the cuprates' phase diagram. We simply find no experimental evidence for the specific \cite{Varma} pattern of circulating currents in the studied compounds according to the theoretical proposal and in contrast to the experimental observation of the effect in pristine Bi2212 \cite{Kaminski}. One should notice, however, that the reason for non-observation of the effect can in our case be the domains of the size smaller than 300 $\mu$m. While the present results put strong constraints on the possible scenarios involving the circulating currents, in some particular cases the obtained information is not sufficient to make a conclusive statement. For instance, while antiferromagnetic pattern of currents required by the d-density wave instability \cite{Chakravarty} is invariant upon reflections in the $\Gamma$ - ($\pi$, 0) plane, reflection in the $\Gamma$ - ($\pi$, $\pi$) plane results in a pattern of currents which can be obtained shifting the initial one by one unit cell vector (i.e. this plane is a glide plane) and therefore cannot be distinguished from the mirror plane reflection in a photoemission experiment. In this case, perhaps, more detailed knowledge as for the energy and momentum distribution of the dichroic signal is required.


We are grateful to C. Varma for numerous stimulating discussions and to R. H\"ubel for technical support. We acknowledge the support of the European Community - Access to Research Infrastructure action of the Improving Human Potential Programme. HB is grateful to the Fonds National Suisse de la Recherche Scientifique.


\begin{thebibliography}{}
\bibitem{spot} This has been checked for two perpendicular directions, rotating the analyzer by 90$^{\circ}$.
\bibitem{remark} We point out that since a strong dichroic signal was observed outside the mirror plane also in the normal incidence geometry (also by us) it can not originate from the non-coplanarity of these three vectors as stated in Ref. \onlinecite{Kaminski}.
\bibitem{gap} We have found experimental evidence for the anisotropic normal state pseudogap in this sample.
\end{thebibliography}
\end{document}